\documentclass{article}
\begin{document}
\author{V. I.  DROSDOV \thanks{Poste restante St-Peterburg-295, 194295, Russia,E-mail:  GalinaD@astrosoft.ru}}
\title{UNIFIED THEORY OF FIELD WITH MODUL OF  SQUARED CURVATURE AS LAGRANGIAN}
\date{20.03.01}
\maketitle
\begin{abstract}The 4-D theory with connection components~
$\Gamma^k_{mn}$as field variables and module of squared curvature~
$\mid R^k_{lmn}R^{lmn}_k\mid$as Lagrangian is described.The Maxwell equations,
the Lorentz condition and the gravity field equation, that agrees with Newton's theory,
result from equations of motion.
\end{abstract}
\begin{center}
04.20.Fy \qquad\qquad 04.50.+h
\end{center}
\newpage
\pagenumbering{arabic}
\section{INTRODUCTION}
We shall describe three initial assumptions in that order,as they had appeared in our investigation.\\
{\it ASSUMPTION 1.} If we assume that, according to general relativity theory ~(GRT) around a mass point the space-time is central symmetrically $<<$ compressed $>>$,then we must consider for a charge point field a picture that is also central symmetrical, but orthogonal to previous one. We could imagine only the central symmetrical torsion. It means that by moving from the origin of  quasi-Galilean coordinates $x^0=ct$,$x^1$,$x^2$,$x^3$, where the charge is, along say the axis $x^1$  it must be $\Gamma^3_{21}=-\Gamma^2_{31}$ . But this equation will be Lorentz--invariant only if it is complemented to
$$\leqno(1)\left\{
   \begin{array}{lcl}
\Gamma^1_{23}=\Gamma^2_{31}=\Gamma^3_{12}=-\Gamma^1_{32}=-\Gamma^2_{13}=-\Gamma^3_{21}=A_0 \\
\Gamma^2_{30}=\Gamma^3_{02}=\Gamma^0_{32}=-\Gamma^2_{03}=-\Gamma^3_{20}=-\Gamma^0_{23}=A_1\\......................................................................... \end{array}\right.$$
(the dotted line represents equations that result from previous by circular permutation of indices 1, 2, 3 ).
Indeed ( 1 ) and ( 3 ) define a completely antisymmetrical tensor $\Gamma^{kmn}$ and  $A_s$ is the dual to it covector.  $A_s$  seems to be alike an electromagnetic 4-potential.\\
{\it ASSUMPTION 2.}Conditions for connection of a gravity field. choose we as follows:
$$\leqno(2)\left\{
   \begin{array}{lcl}
\Gamma^0_{00}=\Gamma^1_{10}=\Gamma^1_{01}=\Gamma^2_{20}=\Gamma^2_{02}=\Gamma^3_{30}=\Gamma^3_{03}=\Gamma^0_{11}=\Gamma^0_{22}=\Gamma^0_{33}=f_0 \\
\Gamma^1_{11}=\Gamma^0_{01}=\Gamma^0_{10}=\Gamma^2_{21}=\Gamma^2_{12}=\Gamma^3_{31}=\Gamma^3_{13}=\Gamma^1_{00}=-\Gamma^1_{22}=-\Gamma^1_{33}=f_1 \\............................................................................................................ \end{array}
\right.$$
In quasi-Galilean coordinates the connection ( 2 ):\\
a) is complementary to ( 1 ) with staff of nonzero components $\Gamma^k_{mn}$ \\
b)is symmetrical in lower indices;\\
c)like (1),obeys the low:$\Gamma^k_{mn}=\Gamma^m_{kn}$ if $k=0$ or $m=0$ and $\Gamma^k_{mn}=-\Gamma^m_{kn}$ otherwise Connection ( 2 ) is it of a conformal flat space with metrics
$$g_{mn}=\exp{2f}g^{(0)}_{mn} \leqno(3)$$
,where $g^{(0)}_{mn}$  is  metrics of flat space, so that $ f_k=f_{,k}=\partial f/\partial x^k $ \\
{\it ASSUMPTION 3.}The sum of all  $(R^k_{lmn})^2$ is the most exact indicator of curvature of space.It vanishes only if all $R^k_{lmn}=0$.But it is not a scalar.The quantity  $$R^k_{lmn}R^{lmn}_k=L\leqno(4)$$ is scalar but it is not positive-defined. We choose  $|L|$  as Lagrangian.\\
\section{EXPOSITION OF THE THEORY}
Let $M$ be a 4-D space-time with metrics ( 3 ) and connection,restricted by( 1 )and( 2 ).In~$M$(see
Appendix 1)
$$\leqno (5)
\begin{array}{l}
L=4[g^{kk}(f_{k,k}+f^2_k-A^2_k]^2+8g^{kk}g^{mm}f_{k,k}\times\\
\qquad{}(f^2_m-A^2_m)-4[g^{kk}(A_{k,k}+2f_k A_k)]^2-16g^{kk}\times\\
\qquad{}g^{mm}A_{k,k}f_m A_m-8g^{kk}g^{mm}(A_{k,m}-A_kf_m-A_mf_k)^2+\\
\qquad{}8g^{kk}g^{mm}(f_{k,m}-f_k f_m+A_k A_m)^2
\end{array}$$,
or in covariant form (see Appendix 2):
$$\leqno (6)
\begin{array}{l}
L=8f_{k;m}f^{k;m}-8A_{k;m}A^{k;m}+4(f^k_{;k})^2-4(A^k_{;k})^2+\\
\qquad{}16f_{k;m}(f^kf^m+A^k A^m)-16f^k_{;k}(f_m f^m+A_n A^n)+\\
\qquad{}12(f_k f^k)^2+12(A_k A^k)^2+24f_k f^k A_m A^m
\end{array}$$
 This expression is wonderful symmetric in $f_k$ and $A_k$ what indicates the good choice of conditions ( 2 ).(Note that in general  $R^k_{lmn}R^{lmn}_k$ includes
5280 summands (the sum of  96 different squared components $R^k_{lmn}$ with 55 summands in each square).It is simple as  (6) only if the triad  ( 1 ), ( 2 ), ( 4 ) is used).\par
By( 2 )and $A_k\equiv 0$;$ k=0, 1, 2, 3$the Ricci scalar
$$R=g^{mn}R^k_{mnk}=6(g^{kk} f_{k,k}+f_m f^m)=6(f^k_{;k}-f_m
f^m)\leqno(7)$$
We shall minimize the integral $$J=\int |L|\sqrt{-g} d^4 x=\int L' d^4 x\leqno(8)$$
The corresponding equations of motion are (see Appendix 3):$$\leqno(9)\left\{
   \begin{array}{lcl}
-24(f^m_{;m}-f_m f^m)_{;k}-48(A^m_{;m}) A_k+24(A_m A^m)_{;k}=0\\
16(A^{k;m}-A^{m;k})_{;m}-48(f^m_{;m}-f_mf^m)\times\\
A^k+24(A^m_{;m})^{;k}+48(A_m A^m)A^k=0\\
\end{array}
\right.
$$
If we suppose that $A_m$ is in proportion to electromagnetic 4-potential with small coefficient and don't take into account $A_m A^m$ as infinitisimals, then we obtain from (9) at once the Maxwell equations in vacuum  $$(A^{k;m}-
A^{m;k})_{;m}=0\leqno(10)$$
the Lorentz condition  $$A^m_{;m}=0\leqno(11)$$  and, as we suppose, the equation
of gravitation field in vacuum  $$f^m_{;m}-f_m f^m=0.\leqno(12)$$
The central symmetrical solution of (12) is $$f=\ln(1-\gamma m/c^2 r)\approx -
\gamma m/c^2 r,\leqno(13)$$
where$\gamma$ is the gravity constant and $m$ is mass.\par
The equations ( 12 ), ( 7 ) and ( 2 ) formally lead to Nordstrom's theory of gravitation,
that affirms:\\
a)geodesics of nonzero length are conics in accordance with  Newton's theory;\\
b) gravitational displacement of spectral lines is the same as in GRT
( the other two consequences see lower).\\
{\it EXPERIMENTALLY} this theory can be proofed by detecting the light polarization
rotation in electromagnetic field in vacuum in accordance with~ ( 1 ).\\
 \section {DIFFICULTIES OF THE THEORY.}
In addition to ( a ) and ( b), Nordstrom's theory affirms:\\
c)displacement of planet perihelion is less then it in GRT and has the other sign;\\
d) light rays don't bend in gravitation fields.\\
It seems to be at variance with experiments, therefore Nordstrom's theory was
rejected. But now it proves to be tight connected with electromagnetic theory.
This advantage may be realized if those two effects will be explained without geometry.\\
{\it ACKNOWLEDGEMENT.} I thank academician L.D.Faddeev and professor
V.A.Franke for helpful discussions and advices.
\section*{APPENDIX 1.}Conditions ( 1 ) and ( 2 ) distribute the nonzero components
$R^k_{lmn}$ to seven groups (we don't write components,result by permutation of
the last indices, but multiply the sum by 2):\\
$$R^k_{klm}=\Gamma^k_{km,l}-\Gamma^k_{kl,m}+\Gamma^k_{pl}\Gamma^p_{km}-
\Gamma^k_{pm}\Gamma^p_{kl}=4(f_{m,l}-f_{l,m})=0;\leqno(A)$$

$$R^0_{110}=R^1_{010}=f_{1,1}-f_{0,0}-A^2_2-A^2_3+f^2_2+f^2_3;\leqno(B)$$
$$........................................................................$$
$$R^2_{332}=R^3_{223}=f_{2,2}+f_{3,3}-A^2_1+A^2_0+f^2_1-f^2_0;$$
$$........................................................................$$
$$R^0_{123}=R^1_{023}=A_{3,3}+A_{2,2}+2(f_1 A_1-f_0 A_0);\leqno(C)$$
$$........................................................................$$
$$R^3_{201}=R^2_{310}=A_{1,1}-A_{0,0}+2(f_3 A_3+f_2 A_2);$$
$$........................................................................$$
$$R^0_{302}=R^3_{002}=(A_{1,0}-A_1 f_0-A_0 f_1)-(f_{3,2}-f_3 f_2+A_3 A_2);\leqno(D)$$
$$........................................................................$$
$$R^0_{203}=R^2_{003}=-(A_{1,0}-A_1 f_0-A_0 f_1)-(f_{2,3}-f_2 f_3+A_2 A_3);$$
$$........................................................................$$
$$R^1_{213}=R^2_{131}=(A_{0,1}-A_0 f_1-A_1 f_0)-(f_{2,3}-f_2 f_3+A_2 A_3);\leqno(E)$$
$$........................................................................$$
$$R^1_{312}=R^3_{121}=-(A_{0,1}-A_0 f_1-A_1 f_0)-(f_{3,2}-f_3 f_2+A_3 A_2);$$
$$........................................................................$$
$$R^3_{031}=R^0_{331}=-(A_{2,3}-A_2 f_3-A_3 f_2)-(f_{0,1}-f_0 f_1+A_0 A_1);\leqno(F)
$$
$$........................................................................$$
$$R^3_{130}=R^1_{303}=(A_{2,3}-A_2 f_3-A_3 f_2)-(f_{1,0}-f_1 f_0+A_1 A_0);$$
$$........................................................................$$
$$R^2_{120}=R^1_{202}=-(A_{3,2}-A_3 f_2-A_2 f_3)-(f_{1,0}-f_1 f_0+A_1 A_0);\leqno(G)
$$
$$........................................................................$$
$$R^2_{021}=R^0_{221}=(A_{3,2}-A_3 f_2-A_2 f_3)-(f_{0,1}-f_0 f_1+A_0 A_1).$$
$$........................................................................$$
By squaring, multiplying by  $\exp {-4f}$  or  $-\exp{-4f}$      depending on is the
number of zeros among indices  $k$,$l$,$m$,$n$ even or odd, and adding, we obtain~ ( 5 ).
(The first, the second and part of the sixth summand in ~( 5) result from group~ (B), the
next  two and part of the fifth --- from group~ (C) and all other --- from all groups
together.)
\section*{APPENDIX 2.}
a)$$f^k_{;k}=f^k_{,k}+\Gamma^k_{kp} f^p=(g^{kk} f_k)_{,k}+4f_p
f^p=g^{kk}f_{k,k}+2f_m f^m\eqno(14)$$
Therefore the first and the second summands in ( 5 ) in sum are:
$$4(f^k_{;k})^2-16f^k_{;k}A_n A^n-12(f_k f^k)^2+4(A_k A^k)^2+24(f_k
f^k)(A_n A^n).$$
b)  Similarly $$A^k_{;k}=g^{kk}A_{k,k}+2f_m A^m\eqno(15)$$
By $k\not=m$   $$A_{k,m}-A_k f_m-A_m f_k=A_{k,m}-
\Gamma^p_{km}A_p=A_{k;m}$$
By $k=m$
$$\begin{array}{l}
(g^{kk})^2(A_{k,k}-2f_k A_k)^2=(g^{kk})^2(A_{k;k}+\Gamma^p_{kk}A_p-2f_k A_k)^2=\\
\qquad{}=\exp {-4f}\{[A_{0;0}-(f_0 A_0-f_1 A_1-\cdots-f_3 A_3)]^2+[A_{1;1}+(f_0 A_0\\
\qquad{}-f_1 A_1-\cdots-f_3 A_3)]^2+\cdots\}=A_{k;k}A^{k;k}-2A^k_{;k}f_m A^m+4(f_m A^m)^2
\end{array}$$
so that  $$-8g^{kk}g^{mm}(A_{k,m}-A_k f_m-A_m f_k)^2=-
8A_{k;m}A^{k;m}+16A^k_{;k} f_m A^m-32(f_m A^m)^2$$
and the third, the forth and the fifth summands in ( 5 ) in sum are $$-4(A^k_{;k})^2-
8A_{k;m} A^{k;m}.$$
e) Since for example  $$f_{0;1}=f_{0,1}-\Gamma^p_{01}f_p=f_{0,1}-2f_0 f_1-
A_3 f_2+A_2 f_3;$$   $$f_{1;0}=f_{1,0}-2f_1 f_0+A_3 f_2-A_2 f_3$$
etc, then
 $$\begin{array}{l}
(f_{0,1}-f_0 f_1+A_0 A_1)^2+(f_{1,0}-f_1 f_0+A_1 A_0)^2=\\
\qquad{}(f_{0;1}+f_0 f_1+A_0 A_1)^2+(f_{1;0}+f_1 f_0+A_1 A_0)^2+2(A_3 f_2-A_2 f_3)^2-\\
\qquad{}2(A_3 f_2-A_2 f_3)(f_{1;0}-f_{0;1})=(f_{0;1}+f_0 f_1+A_0 A_1)^2+\\
\qquad{}(f_{1;0}+f_1 f_0+A_1 A_0)^2-2(A_3 f_2-A_2 f_3)^2
\end{array}$$
etc and by  $k\not=m$
$$\begin{array}{l}
8\sum_{k\not=m} g^{kk}g^{mm}(f_{k,m}-f_k f_m+A_k A_m)^2=8[\sum_{k\not=m} g^{kk}g^{mm}\times\\
\qquad{}(f_{k;m}+f_k f_m+A_k A_m)^2]+8g^{kk}g^{mm}(A_k f_m-A_m f_k)^2.
\end{array}$$
The last sum is $$16f_k f^k A_m A^m-16(f_m A^m)^2.$$
Its sign is plus because$g^{22}g^{33}=-g^{00}g^{11}$etc.
By  $k=m$
$$\begin{array}{l}
(g^{kk})^2(f_{k,k}-f^2_k+A^2_k)^2=(g^{kk})^2(f_{k;k}+\Gamma^p_{kk}f_p-\\
f^2_k+A^2_k)^2=\exp{-4f}\{[f_{0;0}+f^2_0+A^2_0-(f^2_0-f^2_1-\cdots-f^2_3)]^2+\\
\qquad{} [f_{1;1}+f^2_1+A^2_1+(f^2_0-f^2_1-\cdots-f^2_3)]^2+\cdots \} ,
\end{array}$$
therefore
$$\begin{array}{l}
8(g^{kk})^2(f_{k,k}-f^2_k+A^2_k)^2=8(f^k_{;k}+f_m f^m+A^n A_n)^2-\\
\qquad{}16(f^k_{;k}+f_m f^m+A_n A^n)f_p f^p+32(f_m f^m)^2
\end{array}$$
and the last summand in ( 5 ) is equal
$$\begin{array}{l}
8f_{k;m}f^{k;m}+16f_{k;m}(f^k f^m+A^k A^m)-\\
\qquad{}16f^k_{;k}f_m f^m+24(f_k f^k)^2+8(A_k A^k)^2,
\end{array}$$
so that Lagrangian has value ( 6 ).
\section*{APPENDIX 3.}
Let at first $|L|=L$. Since  $\sqrt{-g}=\exp{4f}$,    we can obtain
$L'$  by substitution in ( 5 ) $g_{mn}$ for $g^{(0)}_{mn}$. Therefore
$$\leqno a)
   \begin{array}{l}
\partial L'/\partial f_k=g^{(0)kk}g^{(0)mm}(32f_{m,m}f_k+48f^2_m f_k-\\
\qquad{}48A^2_m f_k-32A_{m,m}f_k-96f_m A_m A_k+16A_{k,m}A_m+16A_{m,k}A_m-32f_{k,m}f_m);
\end{array}$$
$$\leqno b)
\begin{array}{l}
(\partial L'/\partial f_{k,m})_{,m}=g^{(0)kk}g^{(0)mm}[(8f_{m,m}+16f^2_m-16A^2_m)_{,k}\\
\qquad{}+16(f_{k,m}-f_k f_m+A_k A_m)_{,m}]=g^{(0)kk}g^{(0)mm}(24f_{m,m,k}+16f_m f_{m,k}-\\
\qquad{}32A_{m,k}A_m+16A_{k,m}A_m-16f_{m,m}f_k+16A_{m,m}A_k);
\end{array}$$
$$\leqno c)
\begin{array}{l}
\partial L'/\partial A_k=g^{(0)kk}g^{(0)mm}(-32f_{m,m}A_k-48f^2_m A_k+48A^2_m A_k-\\
\qquad{}32A_{m,m}f_k-96f_m A_m f_k+16A_{k,m}f_m+16A_{m,k}f_m+32f_{k,m}A_m);
\end{array}$$
$$\leqno d)
\begin{array}{l}
(\partial L'/\partial A_{k,m})_{,m}=g^{(0)kk}g^{(0)mm}[(-8A_{m,m}-32f_m A_m)_{,k}-\\
\qquad{}16(A_{k,m}-A_k f_m-A_m f_k)_{,m}]=g^{(0)kk}g^{(0)mm}(-8A_{m,m,k}-16A_{k,m,m}-\\
\qquad{}16f_{m,k}A_m-32A_{m,k}f_m+16A_{k,m}f_m+16A_k f_{m,m}+16f_k A_{m,m})
\end{array}$$
and equations of motion are (we multiply them by $e^{-2f}$ or  $e^{-4f}$):
$$\eqno( 16 )\left\{
   \begin{array}{lcl}
g^{mm}[-24(f_{m,m}+f^2_m)_{,k}+48(f_{m,m}+f^2_m)f_k-\\
48(A_{m,m}+2f_m A_m)A_k+48(A_{m,k}A_m-A^2_m f_k)]=0;\\
g^{kk}g^{mm}[16(A_{k,m}-A_{m,k})_{,m}-48(f_{m,m}+f^2_m)A_k+\\
24(A_{m,m}+2f_m A_m)_{,k}-48(A_{m,m}+2A_m f_m)f_k+48A^2_m A_k]=0 \end{array}
 \right.$$
According to ( 14 ), ( 15 ), to $$g^{mm}(W_{mm})_{,k}-2f_k
g^{mm}W_{mm}=(g^{mm}W_{mm})_{,k},$$
where $W_{mm}$   is any tensor, and to  $$g^{kk}g^{mm}(A_{k;m}-
A_{m;k})_{,m}=(A^{k;m}-A^{m;k})_{,m}+4f_m(A^{k;m}-A^{m;k})=(A^{k;m}-
A^{m;k})_{;m}$$
we obtain from ( 16 ) the equations ( 9 ). If $|L|=-L$ the equations of motion and
all other remain the same.\\
{} \\
{} \\
\bf\footnotesize{Drosdov  Vladimir  Ivanovich,
For correspondans: Russia, 194295,    -St-Peterburg-295, post restant Drosdov V. I.
Address: Russia, 194295, St-Petersburg - 295,  Pr.Khudojnikov, 27-1-238.
                          Tel. 599-61-98,}

\end{document}